\documentclass[aps,prd,twocolumn,showpacs,superscriptaddress,groupedaddress,nofootinbib]{revtex4}  % for review and submission

\usepackage{graphicx}% Include figure files
\usepackage{dcolumn}% Align table columns on decimal point
\usepackage{bm}% bold math
\usepackage{amssymb}   % for math
\usepackage[english]{babel}
\usepackage{german}
\usepackage{physics}
\usepackage{csquotes}
\usepackage[export]{adjustbox}
\usepackage{setspace}
\newcommand{\cmmnt}[1]{\ignorespaces}
\usepackage{color}

\usepackage[caption=false]{subfig}

\begin{document}

%\preprint{APS/123-QED}

\title{Self-consistent construction of virialized wave dark matter halos}
% Force line breaks with \\

\author{Shan-Chang Lin}
	\affiliation{%
Department of Physics, National Taiwan University, Taipei 10617, Taiwan
}%
 %\altaffiliation{Physics Department, National Taiwan University.}%Lines break automatically or can be forced with \\
\author{Hsi-Yu Schive}
 \affiliation{National Center for Supercomputing Applications, University of Illinois, Urbana-Champaign, IL, 61820, USA}%
\author{Shing-Kwong Wong}
 \affiliation{%
Department of Physics, National Taiwan University, Taipei 10617, Taiwan
}%
\author{Tzihong Chiueh}%
 \email{chiuehth@phys.ntu.edu.tw}
 \affiliation{%
Department of Physics, National Taiwan University, Taipei 10617, Taiwan
}%
 \affiliation{%
Center for Theoretical Physics, National Taiwan University, Taipei 10617, Taiwan
}%


%\affiliation{
%\affiliation{
%\affiliation{%



\begin{abstract}

Wave dark matter ($\psi$DM), which satisfies the Schr\"odinger-Poisson equation, has recently attracted substantial attention as  a possible dark matter candidate.  Numerical simulations have in the past provided a powerful tool to explore this new territory of possibility. Despite their successes to reveal several key features of $\psi$DM, further progress in simulations is limited, in that cosmological simulations so far can only address formation of halos below $\sim 2\times 10^{11} M_\odot$ and substantially more massive halos have become computationally very challenging to obtain.  For this reason, the present work adopts a different approach in assessing massive halos by constructing wave-halo solutions directly from the wave distribution function.  This approach bears certain similarity with the analytical construction of particle-halo (cold dark matter model).  Instead of many collisionless particles, one deals with one single wave that has many non-interacting eigenstates.  The key ingredient in the wave-halo construction is the distribution function of the wave power, and we use several halos produced by structure formation simulations as templates to determine the wave distribution function.  Among different models, we find the fermionic King model presents the best fits and we use it for our wave-halo construction. We have devised an iteration method for constructing the nonlinear halo, and demonstrate its stability by three-dimensional simulations. A Milky-Way-sized halo has also been constructed, and the inner halo is found flatter than the NFW profile. These wave-halos have small-scale interferences both in space and time producing time-dependent granules. While the spatial scale of granules varies little, the correlation time is found to increase with radius by one order of magnitude across the halo.

%\begin{description}
%\end{description}
\end{abstract}

\maketitle

%\tableofcontents
%\setstretch{1.5}%increase line space

\section{\label{sec:level1}Introduction}
In past decades, tensions of sub-galactic scales between standard cold dark matter (CDM) predictions and galaxy observations have attracted much attention. Examples include the cusp-core problem \cite{de_blok_core-cusp_2010} and the too big to fail problem \cite{boylan-kolchin_too_2011}. Dissipationless simulations of CDM structure formation have found cusp-like density profiles in central regions of halos regardless of the halo mass \cite{navarro_structure_1996,moore_resolving_1998}, while observations reveal that the density profiles of the inner-most region in dwarf spheroidal galaxies favor flat cores \cite{moore_evidence_1994,flores_observational_1994,blok_dark_1997,amorisco_core_2012}. %The predicted subhalos of CDM \cite{klypin_where_1999} outnumber the satellite galaxies detected in the Local Group \cite{mateo_dwarf_1998}.  These satellite galaxies may have lost most of their stars through tidal interactions with host galaxies, rendering them too dim to be detected.
In addition, the most massive subhalos of Milky-Way-sized halos presented by CDM simulations are too massive to account for the observed Milky Way's satellites, dubbed as the too-big-to-fail problem. While these issues may be caused by the limitation of the survey methods, the sensitivity of observations or some not fully explored astrophysics such as baryonic feedback that removes the stars \cite{bullock2017small}, these problems can be signs of trouble against the CDM model, despite the fact that it works well on much larger scales.

Alternative dark matter models have been proposed to solve some of these small-scale problems. One example of these models is the scalar field dark matter (SFDM). It can be divided into two categories, with \cite{goodman_repulsive_2000,guzman_scalar_2000} or without \cite{hu_fuzzy_2000,sin_late-time_1994,schive_cosmic_2014,marsh2016axion,hui2017ultralight,zhang2017evolution,zhang2017cosmological,mocz2018schrodinger} self-interactions. The model without self-interactions, called the wave dark matter ($\psi$DM) or the fuzzy dark matter (FDM), is unique and novel, exploiting the difference in wave and particles dynamics on small scales while keeping large scales identical. Particles of $\psi$DM are non-relativistic extremely light bosons of mass around $10^{-22}$eV, where wave effects, such as interference, appear on astrophysical scales. Since the particle mass is so light that the critical temperature of forming Bose-Einstein condensation (BEC) exceeds the Planck scale, the $\psi$DM is strongly in the BEC state with an infinite phase coherence length and all bosons share the same wave function.
The origin of these extremely light bosons may arise from axions in the string theory \cite{arvanitaki2010string,svrcek2006axions} or a non-QCD axion mechanism in the dark sector \cite{chiueh2014dark}. 
The uncertainty principle renders $\psi$DM to avoid the central cusp formation and helps suppress small-scale structures such as satellite galaxies. On the large scales, $\psi$DM behaves like CDM, in agreement with large-scale observations, such as cosmic microwave background (CMB) observations, where the CDM model is extremely successful.

The first high-resolution cosmological simulation of $\psi$DM structure formation was conducted in 2009 \cite{woo_high-resolution_2009}, in which the core problem was still elusive. Not until 2014 the first adaptive-mesh-refinement simulation came along, able to zoom into the central regions of dwarf galaxies \cite{schive_cosmic_2014}, and discovered that the dark matter halo contains a prominent solitonic core. The core is surrounded by an extended halo, which consists of many small-scale density granules. The sizes of the solitonic core and these granules increase with decreasing halo masses, and are about kpc for dwarf halos. Though the surrounding halo contains most of the mass, composed by the excited states of BEC, the much less massive ground state, i.e., the solitonic core, contributes to a sizable fraction of the gravitational potential depth in the halo center, and thus is a highly nonlinear object. The core and the halo are found to obey the core mass-halo mass relation \cite{schive_understanding_2014}, a relation derived from a nonlocal uncertainty principle.  

However, $\psi$DM simulations have their own limitations, most notably the inability to cover a large volume while maintaining high spatial resolution at the same time.
Particularly troublesome is in the region with the smooth, low-density infalling matter in the vast cosmic volume that can normally be handled with relatively low resolution in ordinary CDM and hydro simulations.  This low-density matter must be captured with high resolution in wave mechanics simulations to resolve the matter wave oscillation; otherwise the infall velocity will be in large error, seriously affecting the mass accretion rate.  Such a difficulty has been circumvented by simulating a small spatial domain, with the drawback that the total mass in the domain is small and therefore halos so formed are often limited to dwarf galaxies \cite{chan2017stars,schive_cosmic_2014,schive_understanding_2014}.  The present work is motivated by this limitation of wave mechanics simulations and aims to find a procedure to construct a realistic 3D virialized halo of arbitrary mass.  

This paper is organized as follows. In Sec. \ref{sec2}, we provide a foundation to connect the classical particle distribution function and the wave distribution function (DF), and then proceed by analyzing dark matter halos obtained by cosmological simulations with the eigenfunction expansion, assuming dark matter halos are in the steady state and spherical-symmetric. We fit the wave distribution function by several classical distribution function models of self-gravitating collisionless particles \cite{binney_galactic_2011}, and identify the best-fit distribution function to be the fermionic King model \cite{chavanis_coarse-grained_1998}. We develop a novel iteration method to solve self-consistent solutions in Sec. \ref{sec:SelfC}. A series of self-consistent solutions are shown and discussed in Sec. \ref{sec:SCresult}. In Sec. \ref{sec:sta}, we demonstrate the stability of these self-consistent halos via three-dimensional numerical simulations. To understand dynamical properties of granules, we examine the temporal and spatial correlation functions of halo density fluctuations in Sec. \ref{sec:time}. Finally, we conclude in Sec. \ref{sec:conclude}.

\section{Wave distribution function of $\psi$ dark matter\label{sec2}}
The $\psi$DM is described by a wave function, which is a classical field, obeying
the Schr"odinger-Poisson equation:
\begin{align}
i\hbar \frac{\partial\psi}{\partial t}&=-\frac{\hbar^2}{2m}\nabla^2\psi+mV\psi,\label{eq:1}\\
\frac{\nabla^2V}{4\pi G}&=m\left|\psi\right|^2\label{eq:2},
\end{align}
where $\hbar$ is Planck constant, $G$ is Newton's gravitational constant, $V$ denotes the gravitational potential of the mass density $m|\psi|^2$ for BEC, and $m$ is the boson mass. We set $m=8.1\times 10^{-23}$eV/c$^2$ throughout this paper. %Here, we have to subtract background density, $\rho_0=3H^2/8\pi G$, assuming the matter dominated universe, and as we discuss the universe today therefore setting $a=1$.
%The derivation of Schr"odinger-Poisson equation can be found in, for example, \cite{marsh_axion_2015}.

\subsection{Statistical mechanics of a single-particle wave function in dynamical equilibrium}\label{sec:DF}

Here we emphasize that $\psi({\bf x})$ is a classical field, and the analysis to follow is different from many-body quantum mechanics, which addresses many possibly configurations of the wave function. For equilibrium systems, the number density of BEC bosons can be expressed as
\begin{align}
&\left|\psi({\bf x},t)\right|^2\notag\\
&=\left|\sum_ia_i\Phi_i({\bf x})e^{-iE_it/\hbar}\right|^2\\
&=\sum_i\left|a_i\right|^2\left|\Phi_i({\bf x})\right|^2+\sum_{i\neq j}a_ia_j^*\Phi_i({\bf x})\Phi_j^*({\bf x})e^{i(E_j-E_i)t/\hbar}\label{eq:Pro}\\
&=\sum_iN\expval{\hat{\rho}_{ii}}{{\bf x}}+\sum_{i\neq j}\expval{\hat{\delta}_{ij}}{{\bf x}},\label{eq:Den}
\end{align}
where $\Phi_i({\bf x})$ is the $i^{th}$ eigenfunction, $a_i$ is a random complex coefficient and $|a_i|^2$ is the weighting factor, which is proportional to the probability, of the $i^{th}$ state, and N is the total number of bosons. The last equality is to bring out the difference between the many-body quantum mechanical density matrix $\hat{\rho}$ and the number density of the classical field. The first term is identical to the coordinate space representation of the density matrix in a mixed state. $\hat{\rho}_{ii}=\sum_i P_i\ket{\Phi_i}\bra{\Phi_i}$, where $P_i$ is the probability of system in state $\ket{\Phi_i}$. The second term only exists in the classical field that represents the interference of different eigenstates, a time-dependent feature that does not exist in many-body quantum mechanics. 
We denote a $\delta$ matrix for this interference term. The interference plays an essential role in the halo, in that it produces the halo granules and provides pressure support against self-gravity.

When we take short-time average and the random-phase average with respect to $a_i$, the number density $<|\psi({\bf x},t)|^2>$ has only the diagonal terms due to the random phase assumption. That is,
\begin{align}\label{eq:gen}
<|\psi({\bf x},t)|^2>=\sum_i\left|a_i\right|^2\left|\Phi_i({\bf x})\right|^2,
\end{align}
which is independent of time in a steady state. Since $|a_i|^2$ is still a random positive factor, we need to further average over different states to smooth out this random factor. The average can often be provided by the summation over degenerate states. However, to do so we need an ansatz. As $a_i$ is a random complex number, we let $a_i=r_i\sigma_ie^{i\phi_i}$, where $r_i$ is a real random number of unity variance, $\phi_i$ a random phase and $\sigma_i$ the variance, and we have $|a_i|^2=\sigma^2_ir^2_i$. The ansatz is that $\sigma_i$ for every degenerate state $i$ is the same. That is, $\sigma_i=\sigma_I$. Hence
\begin{align}
<|\psi({\bf x},t)|^2>=\sum_I\sigma_I^2\sum_{j\in I}r_j^2|\Phi_j({\bf x})|^2,
\end{align}
where the capital index $I$ refers to eigenvalues and degenerate states have the same $I$. For example, we may take $I=E$. In a spherically symmetric potential, the quantum numbers are $n,l,m$, where $l$ and $m$ are orbital and magnetic quantum number of spherical harmonics and $n$ is the principal quantum number. Eigenstates whose eigenvalues $E_{nl}$ are located within an interval $E-\Delta E<E_{nl}<E+\Delta E$ are degenerate, and $\Phi_j({\bf x})$ refers to those degenerate states having $E_{nl}$ in this energy range. To determine the squared variance $\sigma_E^2$, one can average the positive random factor $|a_i|^2=\sigma_E^2r_i^2$ over the degenerate states, thus giving $\sigma_E^2=<|a_i|^2>_E$.

From Eq.(\ref{eq:Den}) we know $<|\psi({\bf x})|^2>$ equals to the density matrix $N\sum_i\expval{\hat{\rho}_{ii}}{{\bf x}}$, and the density matrix $\hat{\rho}$ satisfies the time-independent von Neumann equation when the system is in equilibrium,
\begin{align}
[H,\hat{\rho}]=0. 
\end{align}
In classical mechanics, we have an analogous equation, the Liouville's equation,
\begin{align}
\frac{\partial F}{\partial t}+\{F,H\}=0,
\end{align}
where $\{\}$ is the Poisson bracket, and $F$ is the phase space distribution function. In the collisionless limit, F is the one-particle distribution function $f$. In equilibrium, $\frac{\partial f}{\partial t}=0$, and the solution is $f(I_c)$, for which $I_c$ is the classical constants of motion. In the short-wavelength (or high quantum number) limit, the average density $<\rho>=N\sum_i\expval{\hat{\rho}_{ii}}{{\bf x}}=\sum_I\sigma_I^2\sum_{j\in I}r_j^2|\Phi_j({\bf x})|^2$ approaches the classical equilibrium density
\begin{align}
\rho_c=\int f(I_c)(\frac{d^3p}{dI_c})dI_c,
\end{align}
where $d^3p$ is the momentum-space volume element. In the same limit, we let the summation $\sum_I\rightarrow\int dI$ and identify $\sigma_I^2\rightarrow f(I_c)$, and the remaining term, the momentum-space volume per invariant $I_c$, can be identified as
\begin{align}\label{eq:app}
\sum_{j\in I}r_j^2\left|\Phi_j({\bf x})\right|^2\rightarrow\frac{d^3p}{dI_c}.
\end{align}
The space dependence on the left-hand side is embedded in the right-hand side, due to the fact that for a given constant of motion $I_c$, such as the energy, a combination of space and momentum, the momentum-space volume element $d^3p$ becomes a function of $I_c$ and $x$. %Figure () compares the right-hand side and left-hand side of Eq. (\ref{eq:app}) for $I_c=E$.

In the Appendix \ref{Wigner}, we offer an alternative reduction of the classical distribution function from wave mechanics using the Wigner function.

\subsection{Procedure for the determination of the wave distribution function}\label{sec:DF1}

With the above fundamentals, we first calculate the density matrix, or the density profile (the first term of Eq.(\ref{eq:Pro})), of an equilibrium $\psi$DM halo with the following procedures. The interference terms in Eq.(\ref{eq:Den}) will be taken care later in the construction of 3D halos. 
%\begin{itemize}

(a) Given an assumed or simulation wave function of a $\psi$DM halo, we compute the shell-averaged gravitational potential $\bar{V}(r)$ assuming spherical symmetry. The typical granules are of small scale but the gravitational potential is of large scale and smooth, so that $\bar{V}$ can be calculated from the density profile using spherical shell average.

(b) Solve eigenvalues and eigenfunctions with the gravitational potential $\bar{V}(r)$ 
\begin{align}\label{eq:5}
\left[-\frac{\hbar^2}{2m}\nabla^2+\bar{V}(r)\right]\Phi=E\Phi.
\end{align}
Thanks to the spherical symmetry of $\bar{V}(r)$, one can adopt separation of variables, $\Phi=R(r)Y^m_l(\theta,\phi)$. Equation (\ref{eq:5}) becomes
\begin{align}
sin\theta\frac{\partial}{\partial \theta}(sin\theta\frac{\partial Y_l^m}{\partial\theta})+\frac{\partial^2Y_l^m}{\partial \phi^2}=-l(l+1)sin^2\theta Y_l^m
\end{align}
and
\begin{align}\label{eq:7}
-\frac{1}{2}\frac{d^2u}{dr^2}+\left(\bar{V}(r)+\frac{l(l+1)}{2r^2}\right)u=Eu,
\end{align}
where $Y_l^m(\theta,\phi)$ is the spherical harmonics with integer $l$ and $m$, and $u(r)\equiv R(r)r$.

(c) Set an upper bound of energy equal to the gravitational potential energy at the virial radius, and solve Eq. (\ref{eq:7}) numerically using LAPACK \cite{laug}.  For a given $l$, one can obtain a series of eigenfunctions and eigenvalues $E_{nl}$ of Eq. (\ref{eq:7}), where a radial quantum number is assigned to the eigenvalue for labeling. Sort these eigenvalues from small to large values and label them from $0$ to $K$, where $K$ is the number of eigenfunctions.  In this way we find a nearly complete set of eigenfunctions for bound states, $\Phi_{nlm}({\bf x})=R_{nl}(r)Y^m_l(\theta,\phi)$, which satisfies Eq. (\ref{eq:5}), subject to the choice of the eigenvalue upper bound associated with the virial radius.
\iffalse
\begin{align}
\left[-\frac{\hbar^2}{2m}\nabla^2+mV(r)\right]\Phi_{nlm}=E_{nl}\Phi_{nlm}
\end{align}
\fi

(d) Decompose the wave function at $t=t_0$ from the simulation data using the eigenfunctions determined above,
\begin{align}
\psi({\bf x},t)=\sum_{nlm} a_{nlm}\Phi_{nlm}({\bf x})e^{-iE_{nl}(t-t_0)/\hbar},
\end{align}
where $a_{nlm}$ is the complex coefficient of the eigenstate $\Phi_{nlm}$. %The element of time-average density matrix is then 
\iffalse
\begin{align}
\bar{\rho}_{nlm;n'l'm'}={a_{nlm}a^{*}_{n'l'm'}}.
\end{align}
We define the distribution function as the diagonal elements of density matrix expanded using the complete orthonormal set $\Phi_{nlm}(r)$
\begin{equation}
\rho_{nlmn'l'm'}=f(E_{nl},l)\delta_{nn'}\delta_{ll'}\delta_{mm'}
\end{equation}
\fi

(e) Following the definition above, calculate the distribution function. For instance, if the DF is only a function of energy, in a certain energy neighborhood it can be expressed as
\begin{align}\label{eq:DF}
f(E)|\Phi_E({\bf x})|^2=\frac{\sum_{\Delta E}|a_{nlm}|^2|\Phi_{nlm}({\bf x})|^2}{g(E)},
\end{align}
where $g(E)$ is the density of states of energy eigenvalues $E_{nl}$ within ($E-\frac{\Delta E}{2}$, $E+\frac{\Delta E}{2}$), and $\Phi_E({\bf x})$ is the amplitude weighted sum of eigenfunctions $\Phi_{nlm}({\bf x})$ in the same energy range near $E$, accounting for the spatial dependence on the right-hand side of Eq. (\ref{eq:DF}). To make the separation of $E$-dependence and ${\bf x}$-dependence more precise, we have
\begin{align}\label{eq:DF1}
f(E)=\frac{\sum_{\Delta E} |a_{nlm}|^2}{g(E)}
\end{align}
and
\begin{align}\label{eq:DFx}
|\Phi_E({\bf x})|^2=\frac{\sum_{\Delta E} |a_{nlm}|^2|\Phi_{nlm}({\bf x})|^2}{\sum_{\Delta E} |a_{nlm}|^2}.
\end{align}
(c.f. Eq. (\ref{eq:rhoav})). It is trivial to show that $\int g(E)f(E)dE=\sum_{nlm}f(E_{nl})g(E_{nl})=M$, the halo mass, from Eqs. (\ref{eq:gen}), (\ref{eq:DF}), (\ref{eq:DF1}) and (\ref{eq:DFx}).

%We can check whether the off-diagonal elements of density matrix vanish by calculating
\iffalse
\begin{align}
\frac{\sum_{\Delta}a_{nlm}a_{n'l'm'}^*}{g(E)}.
\end{align}
\fi
%The result is shown in sec. \ref{sec:PDF}, where we also present the scattering plot of amplitudes in complex plane.
%\end{itemize}

\subsection{Results of wave distribution function}\label{sec:PDF}

We test three different DFs, the King, the Osipkov-Merritt King (OMK), and the fermionic King models to check their fits to simulation halos. The details of the three models are presented in the Appendix \ref{sec:CDF}. We fit the distribution function by minimizing
\begin{equation}
\chi^2=\sum_i\left(\frac{<y_i>-f_M(E_i,l_i)}{\sigma_i}\right)^2
\end{equation}
where $<y_i>$ is the average of squared amplitudes over degenerate states in the $E_i$ and $l_i$ bin, $f_M$ the model DF, and $\sigma_i$ the standard error of the data defined to be
\begin{align}\label{eq:standerr}
\sigma_i=\frac{1}{\sqrt[]{N_i-1}}\sqrt[]{<y_i^2>-<y_i>^2},
\end{align}
which describes the uncertainty of the mean. Here, $E_i$, $l_i$, and $N_i$ are the energy, angular momentum quantum number, and the number of states of the $i^{th}$ bin, respectively. Notice that when fitting DFs only depending on energy $f_M(E)$, we bin data in terms of energy. On the other hand, we bin data in two-dimensional space $(E,l)$ when fitting the OMK model. %Equation (\ref{eq:standerr}) has used the central limit theorem for a random number $y_i$. %Because the amplitude squares will change over time, we expect that the DF obtained by averaging amplitude squares at certain time has uncertainty. From central limit theory, we can calculate the standard error of DF, which is Eq. (\ref{eq:standerr}).

\begin{table}
\caption{Fitting results of Halo A. We have reduced the Qsipkov-Merritt King model to one single variable $Q$ to compute $\chi^2_{red}$. These models are discussed in the Appendix \ref{sec:CDF}. The unit of $\mu$ is [$H_0^2\rho_0Mpc^5h^{-5}m_B$] and the unit of $\beta$ is the inverse of that of $\mu$.}\label{tab:a}
\begin{tabular}[t]{lllll}
\hline 
\\[-.7em]
model&$\chi^2_{red}$&bins
& DoF&parameters\\
\hline
\\[-.7em]
King & 37.81& 60&44&(A,$\beta$,$E_c$)\\
=($1.6\times 10^{-7}$,$11$,$-0.049$)&&&&\\
fermionic King&16.16&60&43&(A,$\beta$,$\mu$,$E_c$)\\
=($2.8\times 10^{-8}$,$17$,$-0.36$,$0.0048$)&&&&\\
%Jaffe & 25.26 & 60 & 46& (a,GM,$E_c$)\\
%Wilson&22.96  &60&46&(A,$\beta$,$E_c$)\\
%Hernquist&25.26&60&46&(a,GM,$E_c$)\\
Osipkov-Merritt King&4.36&5400&46&(A,$\beta$,$r_a$,$E_c$)\\
=($1.0\times 10^{-7},11,0.052,-0.005$)&&&&\\
%Michie\\
\hline
\end{tabular}
\end{table}
\begin{table}
\caption{Fitting results of Halo B.}\label{tab:b}
\begin{tabular}[t]{lllll}
\hline 
\\[-.7em]
model&$\chi^2_{red}$ &bins & DoF&parameters\\
\hline
\\[-.7em]
King & 5.05& 70&40&(A,$\beta$,$E_c$)\\
=($3.6\times 10^{-7}$,$3.0$,$-0.17$)\\
fermionic King&3.91&70&39&(A,$\beta$,$\mu$,$E_c$)\\
=($1.2\times 10^{-7}$, $3.5$,$-1.9$,$0.03$)\\
OMK&5.86&10990&1322&($A,\beta,r_a,E_c$)\\
=($1.1\times 10^{-12},2.2,0.0099,-6.$)\\
\hline
\end{tabular}
\end{table}

The postulate of random phase amplitudes in several energy bins has been tested by examining the amplitudes on the complex plane. The distribution in the bin appears Gaussian-distributed, making it clear that the simulation halo satisfies the random phase assumption.

%\iffalse%%%%%%%%%%%%%%comment%%%%%%%%%%%
\begin{figure*}
\subfloat[]{\includegraphics[width=.3\textwidth]{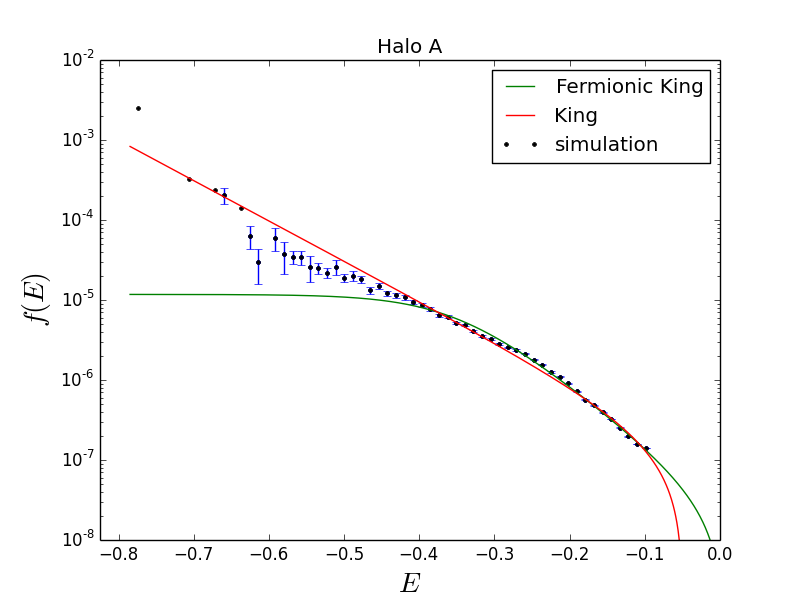}\label{fig:a}}
\subfloat[]{\includegraphics[width=.3\textwidth]{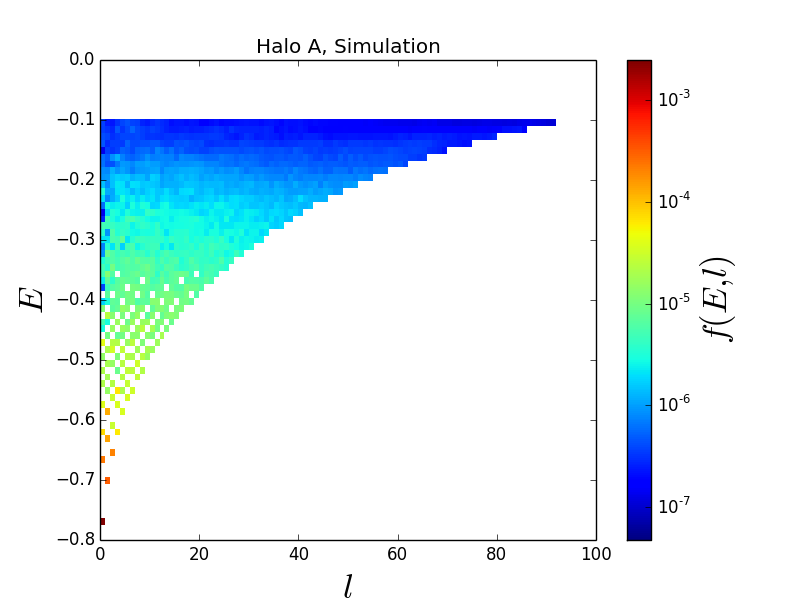}\label{fig:simu}}
\subfloat[]
{\includegraphics[width=.3\textwidth]{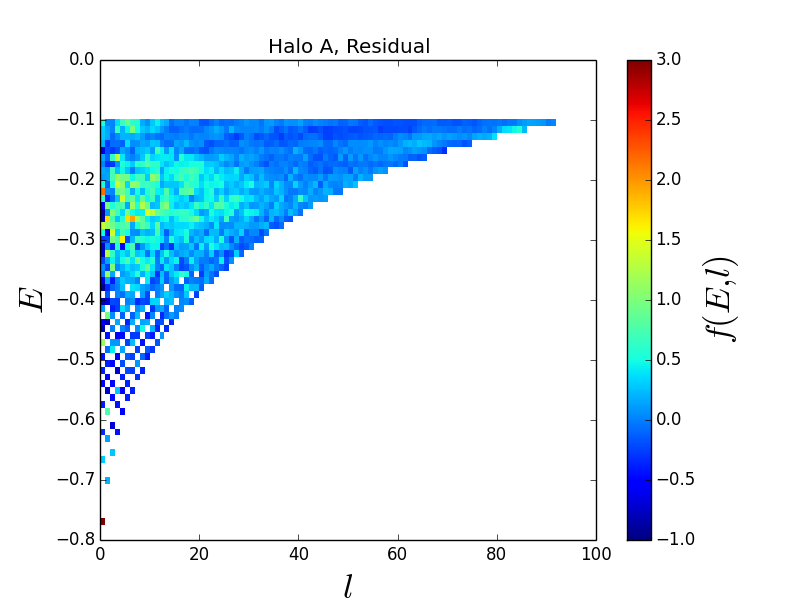}\label{fig:res}}
\caption{ Fitting result of Halo A. (a) Energy distribution functions fitted by the King and the fermionic King models separately.  Error bars represent standard error defined in Eq. (23). While the fermionic King model underestimates the DF in $E \lesssim -0.4$, its $\chi^2_{\rm red}$ is smaller than that of the King model. This is because $\chi_{\rm red}^2$ is dominated by higher-energy bins which have smaller standard errors. If we exclude several higher-energy bins, similar to what we do for Halo B, fermionic King model will follow the DF of lower-energy states. (b) Distribution function in $(l,E)$ space. The lower-right blank region registers no eigenvalue solution. (c) Residual from the best-fit OMK model.}\label{fig:fithaloA}
\end{figure*}
\begin{figure*}
\subfloat[]{\includegraphics[width=.3\textwidth]{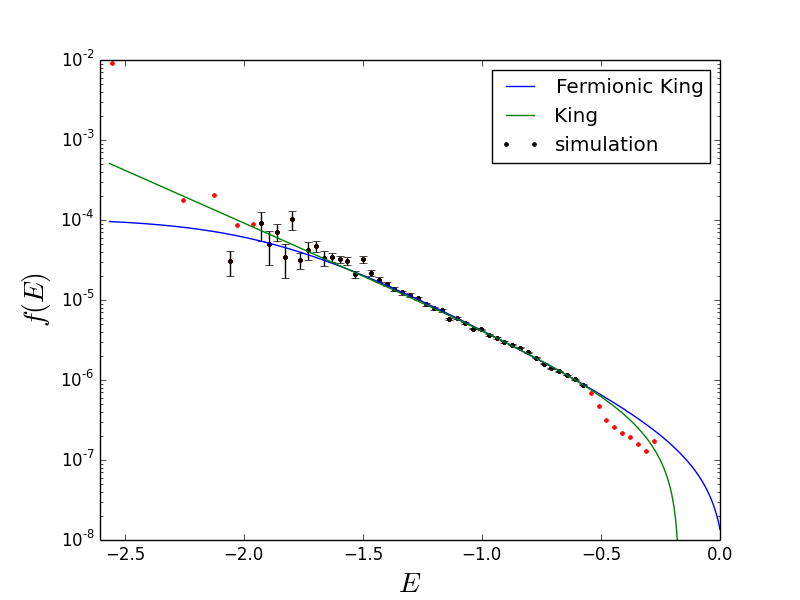}\label{fig:halobDFE}}
\subfloat[]{\includegraphics[width=.3\textwidth]{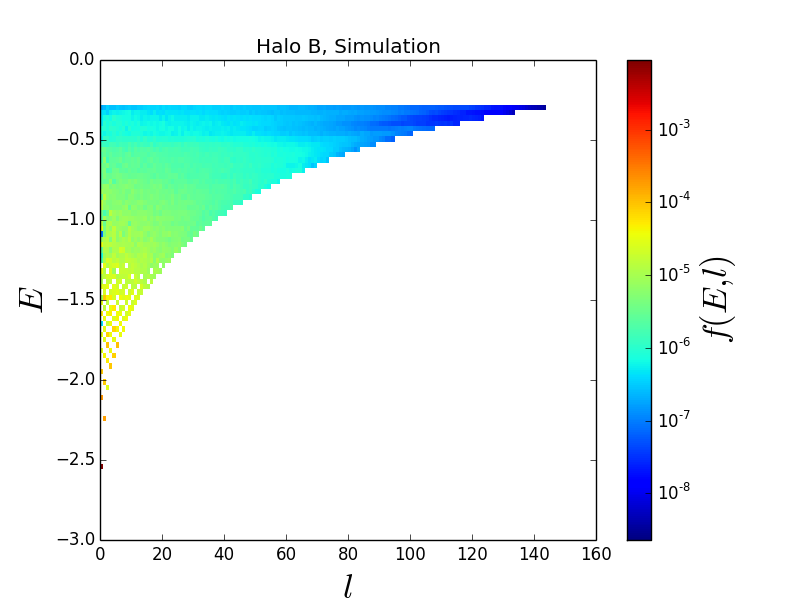}\label{fig:simuB}}
\subfloat[]{\includegraphics[width=.3\textwidth]{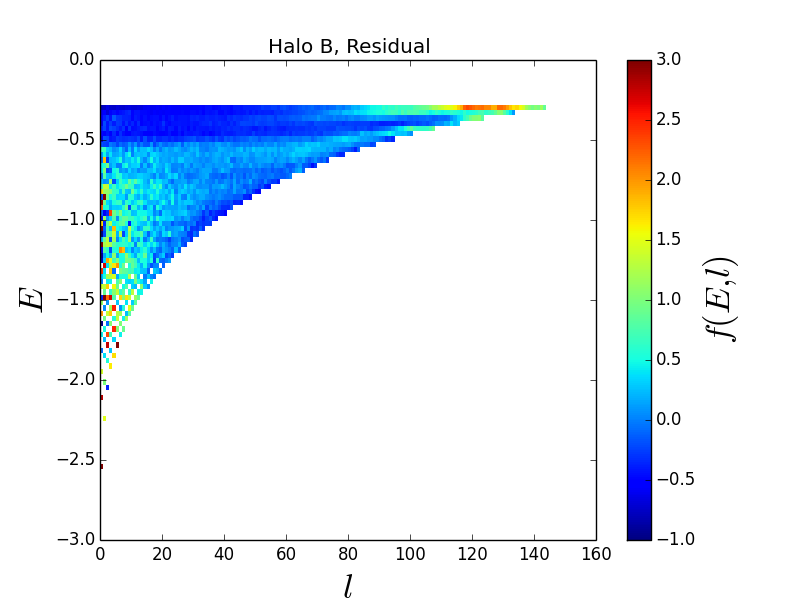}\label{fig:resB}}

\caption{Fitting result of Halo B. (a) Energy distribution functions fitted by the King and the fermionic King models separately.  In the fitting, we ignore the outermost 9 bins shown in red dots. (b) Distribution function calculated in $(l,E)$ space. The lower-right blank region registers no eigenvalue solution. (c) Residual from the best-fit OMK model. We ignore bins whose energy is larger than -0.5.}\label{fig:fithaloB}
\end{figure*}

We analyze five halos, whose masses are $7\times 10^{10}M_\odot$, $2.2\times 10^{10}M_\odot$, $1.7\times 10^{10}M_\odot$, $5\times 10^9M_\odot$, and $2.8\times 10^9 M_\odot$, for the three models. The fitting results of these five halos are similar, and the best-fit reduced chi-square ($\chi_{\rm red}^2$) of two examples of simulation halos, Halo A ($2.2\times 10^{10}M_\odot$) and Halo B ($7\times 10^{10}M_\odot$), are listed in Tables \ref{tab:a} and \ref{tab:b}, respectively. The numbers of eigenstates for Halo A and Halo B are $\sim2.6\times 10^5$ and $\sim 1.0\times 10^6$, respectively. The reduced chi-square is defined as
\begin{align}
\chi^2_{\rm red}=\frac{\chi^2}{\text{degrees of freedom}},
\end{align}
where the degrees of freedom (DoF) equals to the number of bins subtracting the number of model parameters. We use 60 bins for Halo A, and 70 bins for Halo B.

We exclude several high energy bins in some cases. This is due to higher energy modes having dominant contributions in larger radii, and we expect regions near virial radius may not yet reach equilibrium in simulation data. We also exclude the bins with less than five eigenstates due to the large sample variance.  The ground-state bin is also excluded from fitting for the following reason.  The ground state solution produces the soliton which is a highly nonlinear solution, but the probability distribution $f$ is meant to describe the almost interaction-free, excited-state wave functions, analogous to the collisionless particles in classical mechanics. The ground state is hence excluded in the fit of $f$; the amplitude of the ground state solution is instead determined by the soliton mass given by the soliton mass-halo mass relation \cite{schive_understanding_2014}.

\iffalse%%%%
\begin{figure}[h]
\centering
\includegraphics[width=.7\linewidth]{figures/RandomPhases.png}
\captionsetup{width=0.5\textwidth}
\captionsetup{justification=raggedright,singlelinecheck=false}
\caption{Amplitude distribution of eigenvalues between -1 and -0.95. x axis is the real part of amplitudes, and y axis is the imaginary part of amplitudes. The amplitudes distribution is roughly a 2D Gaussian distribution with mean at origin. One can see that random phases is a good assumption. }\label{fig:random}
\end{figure}
\fi%%%%

Figures (\ref{fig:a}) and (\ref{fig:halobDFE}) demonstrate examples of the best-fit results of the King model and the fermionic King model for two simulation halos, Halo A and Halo B. We excluded 9 outermost bins for Halo B when conducting the fitting. One may notice that the fermionic King model underestimates the DF in $E \lesssim -0.4$ for Halo A in Fig. (\ref{fig:a}); however, its $\chi^2_{red}$ is still smaller than that of the King model. This is because $\chi_{\rm red}^2$ is dominated by higher-energy bins which have smaller standard errors. If we exclude several higher-energy bins, similar to what we do for Halo B, the fermionic King model can have a much smaller $\chi^2_{\rm red}$, bringing the model closer to the simulation data.

The fitting results of the OMK model are shown in Figs. (\ref{fig:simu}) and (\ref{fig:res}) for Halo A, and Figs. (\ref{fig:simuB}) and (\ref{fig:resB}) for Halo B. We excluded $E>-0.5$ bins when fitting the OMK model for Halo B. The distribution function $f(E,l)$ is plotted in a two-dimensional color diagram to show the simulation data. The horizontal axis is the orbital angular momentum quantum number $l$ and the vertical axis the energy $E$. The residual is defined as subtracting the simulation distribution function from the best-fit OMK model and then dividing it by the best-fit OMK model. Since the lowest energy eigenvalue increases when $l$ increases, no solution exists in the bottom-right blank region in Fig. (\ref{fig:simu}). It is clear that the simulation data deviate from the OMK model by a large margin for both Halo A and Halo B. The simulation data have prominent low-l components for E around $-0.2$ and $-0.3$ for Halo A and throughout all energies for Halo B, reflecting that the simulation data have strong tangential fringes in the outer halo which will be shown later (in Fig. (\ref{fig:SimuhaloDen})).

Clearly, none of the three models can capture such a prominent low-$l$ feature. Given this fact, one expects to obtain bad fits with the simulation data, and indeed three models have large $\chi^2_{\rm red}$, as tabulated in Tables \ref{tab:a} and \ref{tab:b}. Dominant contributions to $\chi^2$ come from the outer halo where states are highly degenerate and energies are densely packed. With a large number of samples per bin, the error bars are small thereby strongly constraining the models. In all tests, we find the fermionic King model fits better than the other two models, though not significantly better, and therefore from now on the fermionic King model will be adopted for our self-consistent solution construction. We will further demonstrate that halos with the fermionic King distribution is robust and can be very stable in the next section.

\section{Self-consistent solutions of the Schr"odinger-Poisson equation\label{sec:result}}

\subsection{Method of self-consistent solutions}\label{sec:SelfC}

The goal in this section is to solve for all excited-state eigenfunctions that comprise the halo self-consistently.
We have developed a novel iteration method for solving the self-consistent density and potential pair obeying the Schr"odinger-Poisson equations (\ref{eq:1}) and (\ref{eq:2}), assuming that the halo is spherically symmetric. The
self-consistent solution satisfies
\begin{align}
&-\frac{\hbar^2}{2m}\nabla^2\Phi_{nlm}({\bf x})+m\bar{V}(r)\Phi_{nlm}({\bf x})=E_{nl}\Phi_{nlm}({\bf x})\label{eq:27}\\
&\psi({\bf x},t)=\sum_{nlm}a_{nlm}\Phi_{nlm}({\bf x})e^{iE_{nl}t/\hbar}\label{eq:28}\\
%&\psi=\sum_{nlm}a_{nlm}\Phi_{nlm}e^{-i\frac{E_{nl}}{\hbar}({\color{red}t-t_0)}}\label{eq:28}\\
&\nabla^2\bar{V}(r)=4\pi G\bar{\rho}(r)=Gm\int <\left|\psi({\bf x},t)\right|^2>d\Omega\notag\\
&=Gm\sum_{nlm}|a_{nlm}|^2R_{nl}^2(r)\label{eq:29}
%&\approx 4\pi Gm\int^{E_{max}}_{E_0}dEg(E)f(E)|\Phi_E({\bf x})|^2 \label{eq:29}
%&<a_{nlm}a^*_{n'l'm'}>=f_{FK}(E_{nl})\delta_{nn'}\delta_{ll'}\delta_{mm'},\label{eq:30}
\end{align}
where  $\bar{V}({r})$ is the average gravitational potential over solid angle $d\Omega$, $\bar{\rho}(r)$ is the density profile, and $R_{nl}(r)$ are the radial eigenfunctions. Note that we only consider the potential of the average density profiles (c.f., Eq.(7)), and the halo granules are averaged out as they are time-dependent, small compared to the halo size. Note also that the last equality is computationally far less demanding than the second equality that requires full three-dimensional wave functions and suitable for the self-consistent solution search.

For a dark matter halo with mass $M_h$, we make an initial guess with an NFW profile for the halo. The soliton of mass $M_{sol}$ in the core can be specified once the halo mass $M_h$ is given, following the $M_{sol}-M_h$ relation \cite{schive_understanding_2014} and the soliton profile \cite{schive_cosmic_2014}. That is, the initial condition for the iteration is given by
\begin{align}\label{eq:init}
\rho^{(0)}(r)=\Theta(r_e-r)\rho_s(r)+\Theta(r-r_e)\rho_{\rm NFW}(r),
\end{align}
where $r_e$ is the radius at which these two profiles have the same density, $\rho_s(r)$ is the soliton profile, $\rho_{\rm NFW}$ is the NFW profile, and $\Theta$ is the Heaviside step function. The virial radius is defined as the radius within which the average density equals 347 times the critical density given by the spherical collapse model for the $\Lambda$CDM universe. Virial radius is calculated using Eq. (\ref{eq:init}), and it will be fixed in the process of obtaining a halo solution.
%The soliton automatically yields the ground-state solution, and what need to solve for are the excited state solutions.

Given the density, we then compute the corresponding gravitational potential using the Poisson equation, which we call the \enquote{input potential}. Substituting the input potential into the Hamiltonian (Eq. (\ref{eq:27})), we obtain a set of new energy eigenfunctions and eigenvalues. Assign the expectation values for the squared amplitudes of eigenfunctions according to the fermionic King model with given $\beta$ and $\mu$. 
\iffalse
However, we fix the phase of amplitudes to be $\pi/4$, in order to avoid random fluctuations in inner part of halo. More specifically, we set
\begin{align}
&Re(a_{nlm})=Im(a_{nlm})=\sqrt[]{\frac{f_{FK}(E_{nl})}{2}},\\
\text{and therefore}&\\
&|a|^2=f_{FK}
\end{align}
for calculating the time averaged density profile.
\fi
By using these amplitudes, the next step is to construct the wave function utilizing Eq. (\ref{eq:28}), and then we calculate the corresponding gravitational potential by solving the Poisson equation, Eq. (\ref{eq:29}), which we call the \enquote{output potential}. The output potential is generally different from the input potential. If the difference of the input and output potentials is not large, we can adopt a perturbation method. The zeroth order Hamiltonian is
\begin{align}
H_0=-\frac{\hbar^2}{2m}\nabla^2+m\Phi_{\text{in}},
\end{align}
where $\Phi_{in}$ is the input potential.
We take the difference between input and output potentials as the first order perturbation of the Hamiltonian,
\begin{align}
H_1=m(\Phi_{out}-\Phi_{in}).
\end{align}
Perturbation theory demands that the first order correction to the energy is
\begin{align}
\Delta E_{nl}=<nlm|H_1|nlm>,
\end{align}
and the corrected energy
\begin{align}
E'_{nl}=E_{nl}+\Delta E_{nl}.
\end{align}
There is no correction in the eigenfunction to the first order. The new energy is for every eigenstate. The shift in energy changes the expectation values of squared amplitudes according to $f(E)$, and therefore changes the superposed density and the potential. We then update the new Hamiltonian with this new potential to solve eigenfunctions and eigenvalues again. Keep iterating this procedure until the \enquote{input potential} agrees with the \enquote{output potential} to the desired accuracy. We define a dimensionless quantity
\begin{align}
D(V_o,V_i)=\frac{2}{R}\int^R_0\left[\frac{V_o-V_i}{V_o+V_i}\right]^2dr,
\end{align}
where $V_i$ and $V_o$ denotes input and output potential, respectively, and $R$ is the maximum radius for solving the eigenvalue problem Eq. (\ref{eq:27}). We adopt $D(V_o,V_i)<0.01$ as the limiting value for obtaining a self-consistent nonlinear solution satisfying Eqs. (\ref{eq:27}) -- (\ref{eq:29}). 

Note that we fix the ground state amplitude and the halo mass during the iteration. It is worthwhile to point out that although the ground state amplitude is fixed during the iteration, the ground state eigenfunction will change slightly due to the change of the gravitational potential for every iteration. As the ground state is a highly nonlinear object and cannot be described by the distribution function, we set the ground state amplitude from the core-halo mass relation \cite{schive_understanding_2014} although the ground state shape may change during the iteration.

For the iteration method to work, an appropriate initial choice of the input potential is essential. For given parameters $(\beta, \mu)$, one can find an appropriate initial input potential by trial and error. Specifically, we randomly choose a concentration parameter $c$ of the NFW profile in the range $1 \le c \le 30$ until the iteration method mentioned above leads to a converged solution. A larger value of c corresponds to a deeper initial input potential. If the initial input potential is too deep compared with the correct self-consistent potential, % the number of eigenfunctions in the inner part {\color{red} of the halo (outside the core)} will be higher than that of the self-consistent solution as the density of state is approximately proportional to classical phase space volume. For such a potential, 
the output potential would become ever increasingly deeper with iterations, and the solution runs away. A similar situation happens when the initial input potential is too shallow. If the initial input potential is not far from the correct self-consistent potential, the correction of eigenvalues $\Delta E_{nl}$ flips signs at each iteration, and this usually warrants convergence for most $n$ and $l$. Having said that, this perturbative iteration method generally speaking has a relatively large converging radius since the iteration tends to be self-corrective. We find the potential of the NFW density profile often provides a good initial guess. For some parameters $(\beta, \mu)$ of the fermionic King model, the iteration procedure, however, fails to converge no matter what the initial guess is. We consider this case to be the termination of steady-state solution.

Finally having the self-consistent profile, we need to assign random complex amplitude to eigenstates for the three-dimensional halo wave function.  We adopt random complex amplitudes obeying a two-dimensional Gaussian probability distribution. The probability distribution of amplitudes $a=a_r+ia_i$ of a given energy E is
\begin{equation}\label{73}
P(a_r,a_i)=\frac{1}{2\pi\sigma^2}\text{exp}\left(-\frac{a_r^2+a_i^2}{2\sigma^2}\right)
\end{equation}
where
\begin{equation}\label{74}
\sigma=\sqrt[]{\frac{1}{2}f_{FK}(E)}
\end{equation}
and $f_{FK}$ is the fermionic King's distribution.
From Eq. (\ref{73}) and (\ref{74}), we have the average of squared amplitudes $<|a|^2>=f_{FK}(E)$ and ensure the cross term $<a_1a_2^*>=0$ for different eigenstates $1$ and $2$. 

%\multlinecomment{
\iffalse%%comment this paragraph to the line \fi
\subsection{Spin parameter}

Spin parameter, a dimensionless quantity, is a standard parameter which is used to describe total angular momentum of halos produced by simulation. Spin parameter is defined as[]
\begin{align}
\lambda=\frac{|E|^{1/2}L}{GM^{5/2}},
\end{align}
where $M$ is the total mass of the system, $E$ is its total energy and L is its total angular momentum. Spin parameter provide a measurement of angular speed independent of the system's mass. Obviously, $\lambda=0$ for a non-rotating system, and $\lambda=0.4255$ for a cold, razor-thin, self-gravitating exponential disk[]. For comparison, the spin parameter distribution of halos formed in standard CDM cosmology can be fitted by a log-normal distribution with median value $\lambda=0.037$, shown in figure \ref{fig:CDMlambda}.

We have not had a full analysis for spin parameter in $\psi DM$ cosmology due to the lack of simulation runs. However, we present a preliminary study for it, based on the analysis of eigenfunctions decomposition. The spin parameter of $\psi DM$ halos can be calculated as
\begin{align}
\lambda=\sum_{nlm}\frac{|E_{nlm}|^{1/2}m\hbar}{GM^{5/2}}
\end{align}
for z-component, where $m$ is the magnetic quantum number. The result and discussion of spin parameters for the four $\psi DM$ halos we consider will be shown in section three.
%}
\fi

\subsection{Comparison of different self-consistent solutions\protect\footnote{The scaling relation as presented in \cite{schive_understanding_2014} has a factor $2$ scatters; particularly at $z\rightarrow 0$, the soliton mass tends to be higher than the average.  The reason is that in cosmology simulations while the soliton continues to grow in mass, the halo awaits major mergers to grow.  In the presence of a cosmological constant, the waiting time is long compared with the simulation sampling time, thereby yielding a slight deficiency in halo masses of major halos. The simulation halos we adopted as templetes are these late-time halos and the soliton mass
is roughly a factor of 2 higher than the core mass-halo mass relation indicates. We hence increase the core mass defined in \cite{schive_understanding_2014} by 1.7 to the constructed halos so that they can be compared with simulation halos.}}\label{sec:SCresult}

In this section, we will show several examples of self-consistent solutions with different model parameters, and illustrate the effects of changing these parameters. The fermionic King model has four parameters defined in the Appendix \ref{sec:CDF}. The quantity $A$ is fixed for a given halo mass. $E_c$ is the cutoff energy at which the fermionic King model drops to zero, and it is in general larger than the potential energy at the virial radius. In this work, $E_c$ has a negligible impact on DF since the upper limit of energy eigenvalues is smaller than $E_c$. Therefore, we set $E_c$ to zero when constructing self-consistent solutions. The remaining two free parameters are $\beta$ and $\mu$, the inverse temperature and the chemical potential, respectively. The fermionic King model reduces to the King model when $\mu\rightarrow -\infty$. %We also compare the self-consistent density profile with that given by the simulation, and find it agrees with the simulation halo for certain parameters. These parameters are in accord with the distribution function obtained from simulation well. 

\begin{figure}
%\begin{subfigure}{.5\textwidth}
%\centering
\subfloat[]{\includegraphics[width=.5\textwidth]{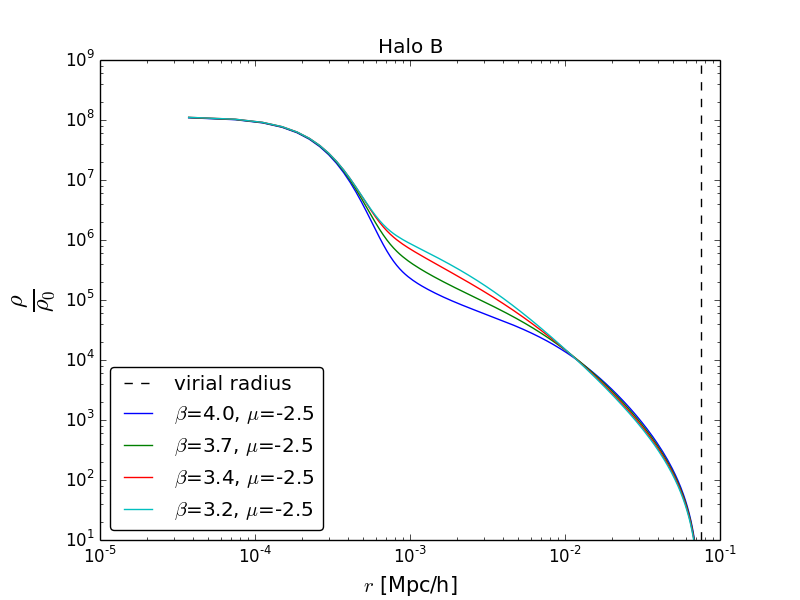}\label{fig:diff_beta}}
\newline
%\caption{Density profiles of self-consistent solutions with different $\beta$.}\label{fig:diff_beta}
\subfloat[]{\includegraphics[width=.5\textwidth]{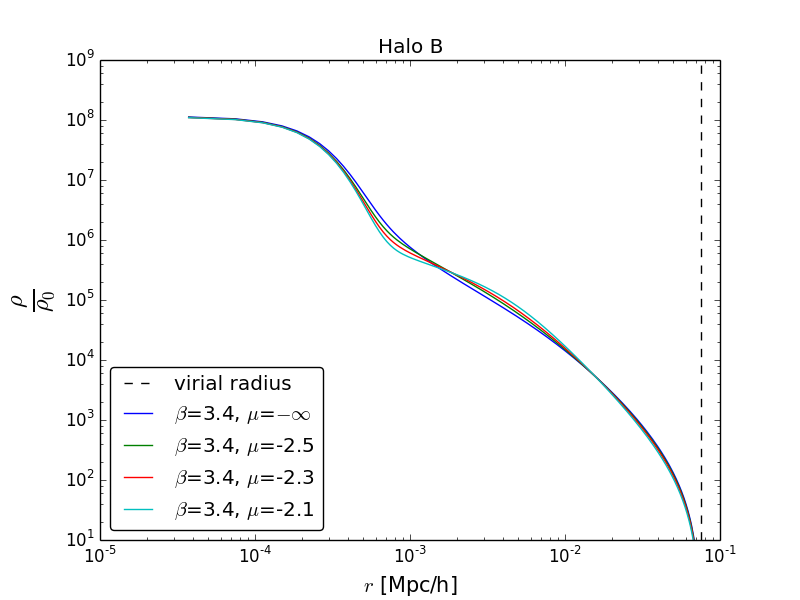}\label{fig:mu}}
\newline
%\caption{Density profiles of self-consistent solutions with different chemical potentials ($\mu$).}\label{fig:mu}
\subfloat[]{\includegraphics[width=.5\textwidth]{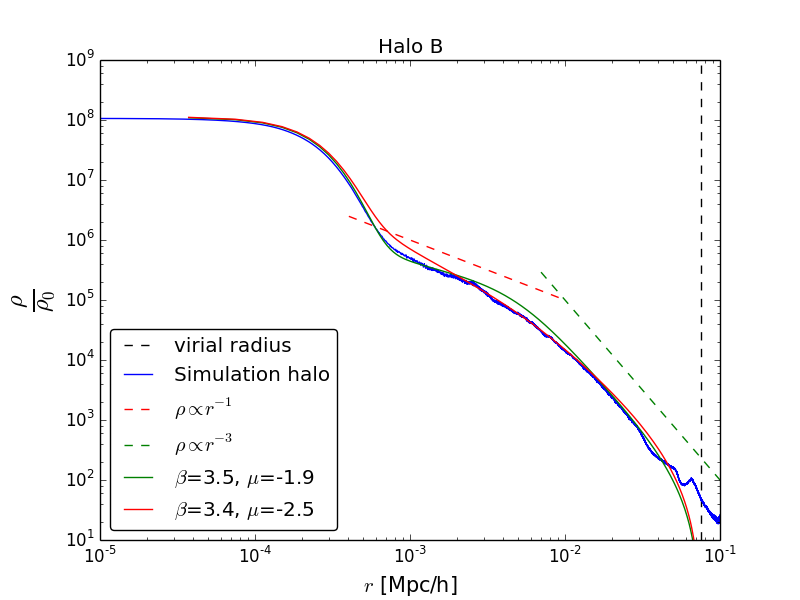}\label{fig:dirfit}}
\caption{(a) Density profiles of self-consistent solutions with different $\beta$. (b) Density profiles of self-consistent solutions with different chemical potentials ($\mu$). (c) Density profiles of self-consistent solutions with parameters ($\beta, \mu$) obtained from (i) the minimum reduced chi-square against the distribution function data (green solid line) and (ii) the closest one to the simulation density profile among Fig. (\ref{fig:diff_beta}) (red solid line). The simulation density profile is also shown (blue solid line). The power laws $r^{-1}$ and $r^{-3}$ are also plotted as references.}
%\end{subfigure}
%\caption{}
\end{figure}

A series of self-consistent halo densities with a given halo mass and $\mu=-2.5$ but different $\beta$ is shown in Fig. \ref{fig:diff_beta}, where $\rho_0$ is the background matter density at present. Moreover, we restrict our discussion to Halo B, which is one of the most massive virialized halos in our cosmological simulations. The mass of this halo is $7\times 10^{10}M_{\odot}$, and note that its core mass is a factor of 1.7 more massive than predicted by the core-halo mass relation explained earlier.

Plotted in Fig. (\ref{fig:diff_beta}) are the density profiles of these constructed halos. The unit of distance is Mpc/h, where h=0.70 is the dimensionless Hubble parameter. As $\beta$ increases (or temperature decreases) the inner halo becomes less concentrated, and the gravitational potential becomes shallower. This feature is expected for a low temperature halo of a fixed halo mass, since the potential must be shallower for a lower temperature virialized system.  On the other hand, Fig. (\ref{fig:mu}) shows halo densities of $\beta=3.4$ but with different chemical potentials $\mu$. We observe that by increasing chemical potential the density becomes flatter in $r \lesssim 3 \times 10^{-3}\rm Mpc/h$ and steeper in $3 \times 10^{-3} {\rm Mpc/h} \lesssim r \lesssim 10^{-2} {\rm Mpc/h}$. This behavior is what one would have expected for the fermionic distribution function where the chemical potential suppresses the amplitudes of eigenmodes when energies are below the chemical potential. These suppressed eigenmodes are those lowly-excited states, and thus contribute to the innermost part of the halo. We surprisingly find that the inverse $\beta$ is higher than the virial temperature of this self-consistent halo roughly by a factor of 3.  This may be related to the slightly non-isothermality of halos elucidated in a later section.

It is important to verify whether the parameters $(\beta, \mu)$ of the self-consistent solutions are in agreement with the DF obtained from the simulation halo (Halo B) when the constructed halo and the simulated halo have almost identical density profiles, and whether the constructed profile with the best-fit ($\beta, \mu$) to the simulated halo reproduces the density profile of Halo B. Among the previous self-consistent solutions, we first identify a self-consistent density profile with $\beta=3.4$ and $\mu=-2.5$ that is the closest to the simulation density profile. We then calculate $\chi^2_{\rm red}$ of the fermionic King model with these values of $\beta$ and $\mu$ against the simulation data and obtain $\chi^2_{\rm red}=5.15$.  This is to be contrasted with the minimum $\chi^2_{\rm red}=3.91$ obtained by fitting DF directly. On the other hand, we also construct the self-consistent solution with the minimum $\chi^2_{\rm red}=3.91$ parameters, $\beta=3.5$ and $\mu=-1.9$. Figure (\ref{fig:dirfit}) shows the density profiles of these two sets of parameters as well as the simulation data. Both profiles fairly resemble, but are not identical to, the simulation density profile. The slight deviation of these solutions from the simulation data reflects the lack of a precise functional form of the distribution function. The checks set an estimate for the limitation of self-consistent solutions. 

\iffalse
\begin{figure}[h]
\includegraphics[width=.5\textwidth]{figures/DF_E_run03halo0203mu-2_5beta3_4.png}
\captionsetup{width=0.5\textwidth}
\captionsetup{justification=raggedright,singlelinecheck=false}
\caption{Distribution function fitted by $\beta=3.4$, $\mu=-2.5$, and $E_c=0$(red) compared with direct fitting (green). Note that we excluded eight outermost bins as it does not seems to be in equilibrium.}\label{fig:DFeye}
\end{figure}
\fi

Figure (\ref{fig:dirfit}) also provides hints about how to determine $\beta$ so that the self-consistent solutions fairly resemble simulation halos. Note that $\beta=3.2$ in Fig. (\ref{fig:diff_beta}) is the lowest value we can construct a self-consistent solution when fixing $\mu=-2.5$. Since the density profile of self-consistent solution with $\beta=3.4$ and $\mu=-2.5$ appears closest to that of simulation, we conjecture that the density profile is similar to simulation when the value of $\beta$ is close to the lowest possible value. This strategy will be adopted for the construction of halos without templates.  On the other hand, so far there is no clear evidence on how to determine $\mu$.

\begin{figure}
\includegraphics[width=\linewidth]{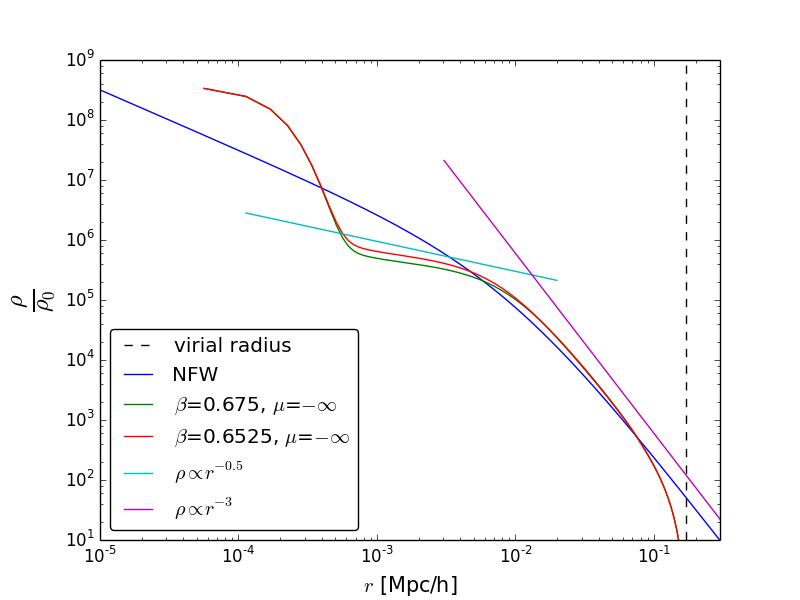}
\caption{Density profiles of self-consistent solutions of a $8\times10^{11} M_{\odot}$ halo. The inner profile and outer profile are close to power laws $r^{-0.5}$ and $r^{-3}$, respectively.}\label{fig:MW}
\end{figure}

The target of this work is to construct the wave function of massive galaxies, with halo mass around $10^{12} M_{\odot}$,
as the current AMR cosmological simulations are incapable of running with a sufficient large volume to form such a massive galaxy \cite{schive_cosmic_2014,schive_understanding_2014}. We have constructed a series of self-consistent solutions for a halo of mass $8\times10^{11}M_{\odot}$, as shown in Fig. (\ref{fig:MW}). %The highest temperature halo has parameters $\beta=0.6525$ and $\mu=-1\times 10^{10}$, 
(For these massive-galaxy halos, we let the soliton mass obey the soliton mass - halo mass relation and have not modified the soliton mass.)
In this case, $\beta = 0.65$ is close to the lowest value, below which no solution can be found.  We find that the inner part of the density profile is already suppressed even for $\mu\rightarrow -\infty$, and therefore this set of parameters are adopted to solve for the Milky-Way-sized halo wave function.  This is basically the King model.  A slightly lower temperature solution is also plotted to show the trend near this solution.  Also plotted in Fig. (\ref{fig:MW}) for references are two logarithmic slopes of $-0.5$ and $-3$ for the inner and outer halo of these solutions, as well as the NFW profile of mass $8\times 10^{11}$ and concentration parameter $c=18$, according to \cite{mcmillan_mass_2011}.

Unlike the previous less massive halos, the inner halo of the massive galaxy is relatively flat compared with the NFW profile of the CDM model, despite the fact that the outer halo appears to be consistent with the NFW profile. We will discuss this difference in Sec. (IV), and three-dimensional simulation tests of these halos will be conducted in a future work.  %If we model the density profile of the halo as $\rho$ 

\subsection{Stability of constructed halos}\label{sec:sta}

\begin{figure}
\subfloat[]{\includegraphics[width=\linewidth]{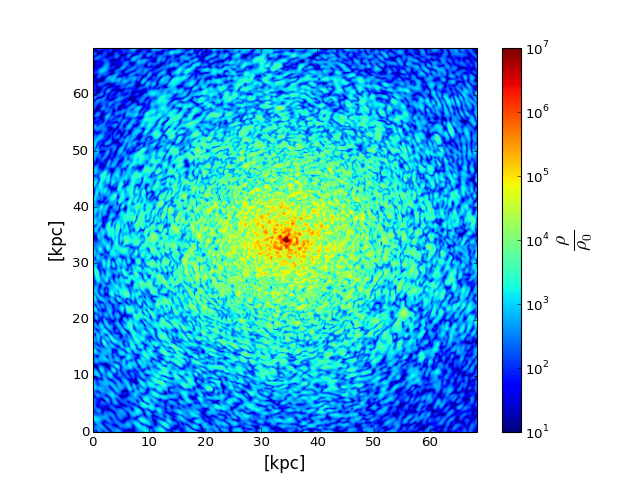}\label{fig:SimuhaloDen}}
%\caption{Density slice of the simulation $\psi$DM halo.}\label{fig:SimuhaloDen}
\newline
\iffalse
\begin{figure}
\centering
\captionsetup{width=\linewidth}
\captionsetup{justification=raggedright,singlelinecheck=false}
\includegraphics[width=\linewidth]{figures/Den_sliceXY400_0_SimPot.png}
\caption{Density slice of the non-self-consistent $\psi$DM halo using simulation potential.}\label{fig:ArtihaloDen1}
\end{figure}
\fi
\subfloat[]{\includegraphics[width=\linewidth]{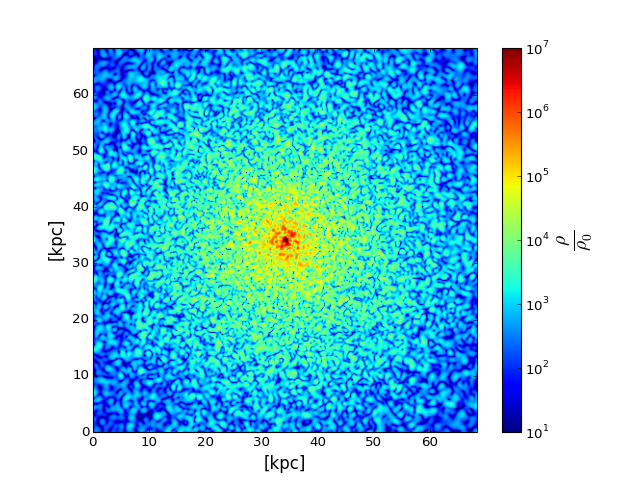}\label{fig:ArtihaloDen2}}
\caption{(a) Density slice of the simulation $\psi$DM halo. (b) Density slice of the self-consistent $\psi$DM halo with $\beta=3.4$, $\mu=-2.5$.}
\end{figure}

Next, we shall examine the stability of the constructed self-consistent halos. We check three dwarf-galaxy-sized halos, (i) simulation halo, (ii) self-consistent halo with $\beta=3.4$ and $\mu=-2.5$, and (iii) self-consistent halo with $\beta=4.0$ and $\mu=-2.5$. Figure (\ref{fig:SimuhaloDen}) shows the slice image of density for the simulation halo cutting through the halo center. Figure (\ref{fig:ArtihaloDen2}) shows the same slice of the self-consistent halo of $\beta=3.4$ and $\mu=-2.5$. The slice of halo (iii) is not shown since it is almost indistinguishable from Fig. (\ref{fig:ArtihaloDen2}). From these slice images, one sees that the self-consistent halo has isotropic granules throughout the halo, whereas the simulation halo has tangentially elongated granules in the outer halo. This image reveals evidence that the distribution function ought to depend on the angular momentum which is not captured by the fermionic King model. 
% artificial halos based on anisotropic distribution function, because we did not find any anisotropic distribution function which fit simulation data well enough compared with isotropic models.
%or it is not virialized in the outer region, while direct fitting faces the challenge that is hard to determine the dependence of angular momentum, which was discussed in section \ref{sec:PDF}.

We demonstrate the stability by evolving the halos for one free-fall time $T_{ff}$ defined as
\begin{align}
T_{ff}=\sqrt[]{\frac{\pi^2r_{vir}^3}{8MG}},
\end{align}
where $r_{vir}$ is the virial radius, M is the enclosed halo mass within the virial radius, and G is the Newton's constant. The boundary conditions are the isolated boundary condition for gravity and rigid-wall boundary condition for wave function. Box size is 180 kpc/h, with 43 pc/h spatial resolution. The simulations are conducted by the GPU-accelerated adaptive mesh refinement code GAMER \cite{schive_gamer:_2010,schive2017gamer}. 

\begin{figure}
%\centering
\includegraphics[width=\linewidth]{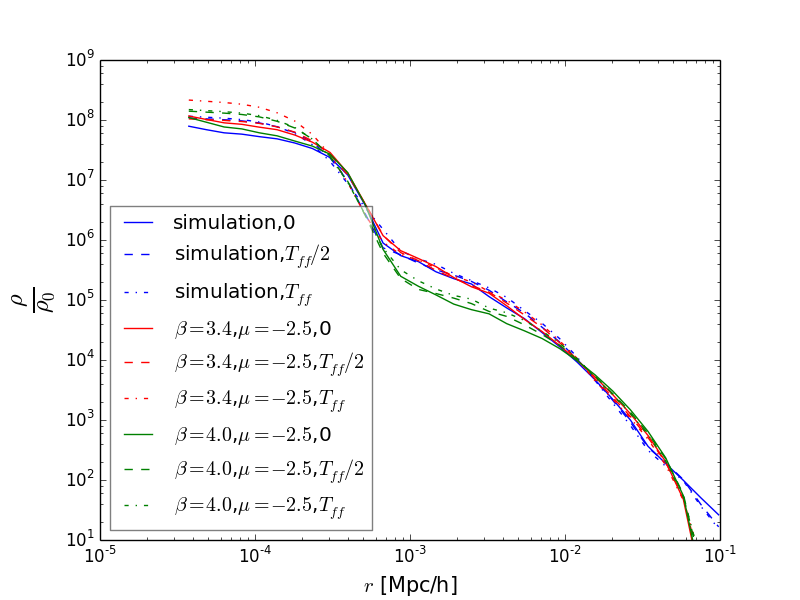}
%\captionsetup{width=\linewidth}
%\captionsetup{justification=raggedright,singlelinecheck=false}
\caption{Density profiles of two self-consistent halos (red and green lines) and simulation halo (blue lines) evolved for one halo free-fall time ($T_{ff}$). It demonstrates that the self-consistent halos are very stable.}\label{fig:evo2}
\end{figure}

Figure \ref{fig:evo2} shows the evolution of these three halos. They evolve roughly in the same manner, with a stable outer halo and a mildly fluctuating soliton. The simulation results demonstrate that our self-consistent halos are stable in the parameter range investigated.  It remains to be seen whether 
a halo of the same mass but with very different parameters is also stable.  But for the purpose of this first work toward the new theoretical approach, we are confined to the parameter regime where we believe to be physical.

\iffalse%%%%%%%%%%%%%%%
\begin{figure}[h]
%\centering
\includegraphics[width=\linewidth]{figures/Fig__Real_vs_Arti__DensProf.png}
%\captionsetup{width=\linewidth}
%\captionsetup{justification=raggedright,singlelinecheck=false}
\caption{Density profiles evolved during one halo free-fall time. Red curves denote the non-self-consistent halo profile constructed by the simulation potential, and blue curves denote the simulation halo profile. The solitons oscillate by a sizable amount in both cases. Moreover, the inner halo of red curve changes substantially while that of blue curve remains stable.}\label{fig:evo1}
\end{figure}
\fi%%%%%%%%%%%%%%%

% We also show the evolution of artificial halo made by soliton plus NFW potential. It relaxes to more stable configuration within a free-fall time from the virial radius.

%\section{Time correlation function}

\begin{figure*}
\subfloat[]{\includegraphics[width=.5\textwidth]{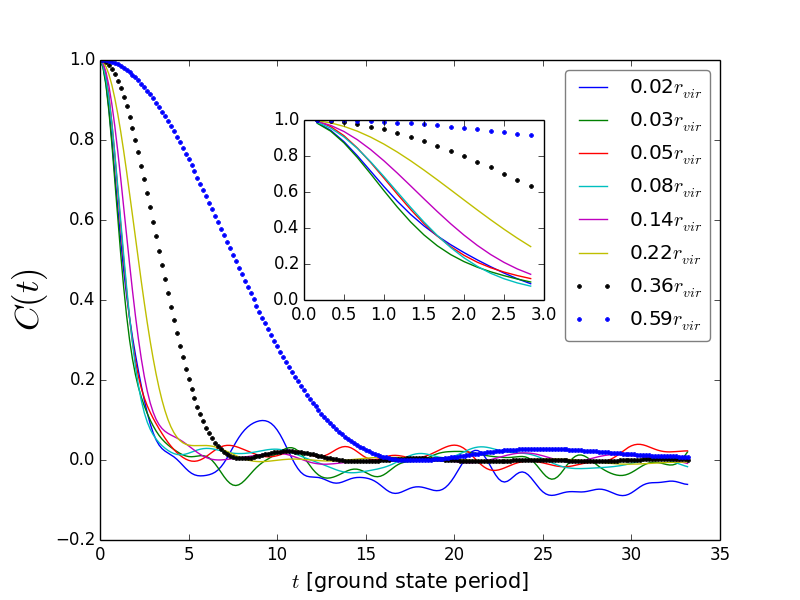}\label{fig:corr}}
\subfloat[]{\includegraphics[width=.5\textwidth]{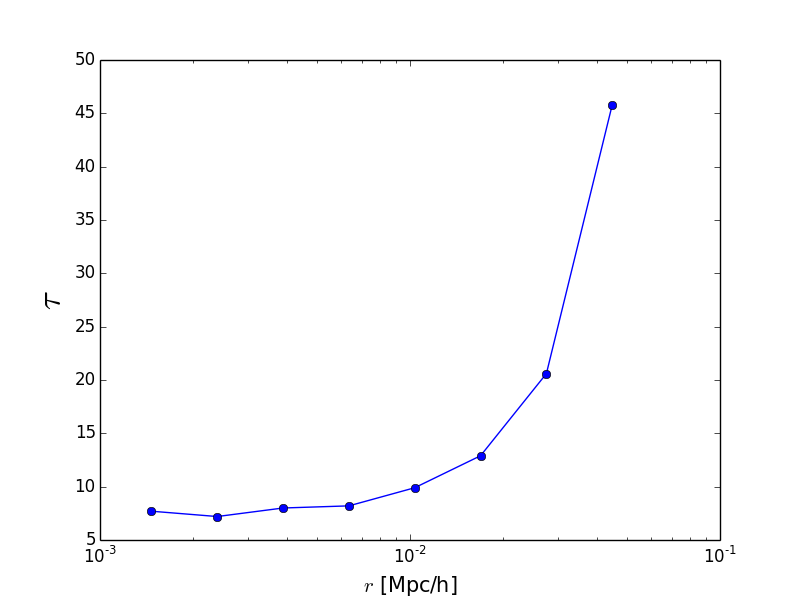}\label{fig:tau}}
\newline
\subfloat[]{\includegraphics[width=.5\textwidth]{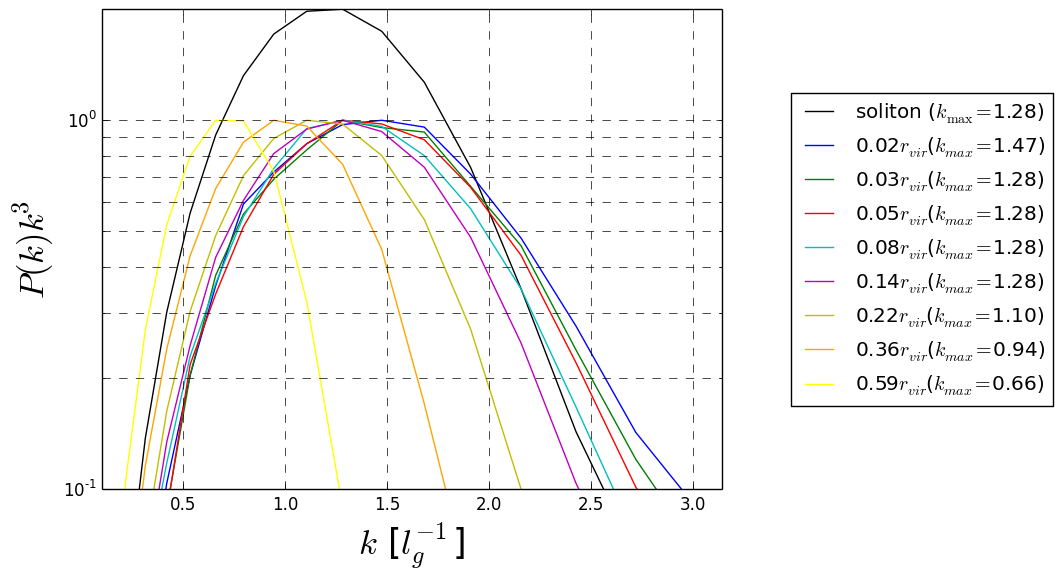}\label{fig:PS}}
\subfloat[]{\includegraphics[width=.5\textwidth]{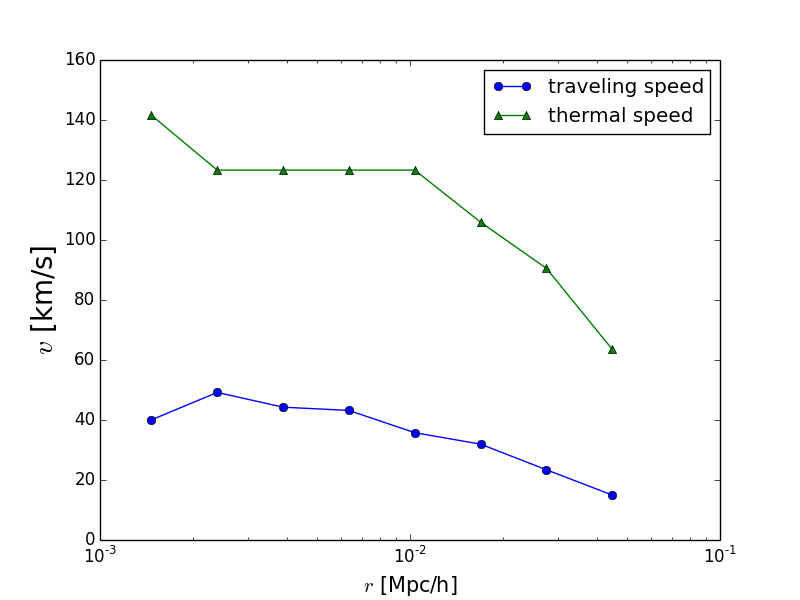}\label{fig:Ther_vs_Trav}}
\caption{(a) Halo time correlation function for the self-consistent solution of $\beta=3.4$ and $\mu=-2.5$ at different radii. The inset shows the correlation function within $0\leq t\leq 3T_g$, where $T_g$ is the ground state period. (b) Correlation time as a function of radius. (c) Power spectrum of halo granules at different radii (same as those in Fig. (\ref{fig:corr})). The unit of $k$ is the inverse of grids length ($l_g=0.17$kpc/h). The typical size of granules increases as radius increases. The amplitudes of granules power spectrum are normalized to one, while that of the soliton is set to a higher value to distinguish the soliton from granules. We find the spectral peaks differ by only a factor of 2 from the innermost radius to the outermost radius. (d) Thermal speed and traveling speed as functions of radius.}
\end{figure*}

\section{Temporal and spatial correlation functions of halo granules}\label{sec:time}

An important issue, which can hardly be addressed by simulations but can be addressed by our theoretical model, is the time dependence of halo granules \cite{chan2017stars}. We introduce the time correlation function for granules as a function of radius, which is defined as 
\begin{align}
C(r_i,\tau)=\int_{r_i-\Delta r/2}^{r_i+\Delta r/2} \delta({\bf r},0)\delta({\bf r},\tau)d^3r
\end{align}
and
\begin{align}\label{eq:delta}
\delta({\bf r},t)=\frac{\rho({\bf r},t)-\bar{\rho}({\bf r})}{\bar{\rho}({\bf r})},
\end{align}
where $\rho({\bf r},t)$ is the density, $\bar{\rho}({\bf r})$ is the average density over a narrow shell, and $r_i$ is the radius of the $i^{th}$ shell with a small width $\Delta r$. The time correlation function measures the granule coherence time at a fixed position.  As granules can die out but will also move around, the coherence time may not be the granule lifetime, but may better be interpreted as the travel time across a granule if the lifetime is longer than the travel time.  When so, we can estimate the granule travel speed if the typical granule sizes as a function of radius are known.

We examine the $7\times 10^{10} M_\odot$ self-consistent halo with $\beta=3.4$ and $\mu=-2.5$. Halo time correlation functions at various radii are shown in Fig. \ref{fig:corr}. %The radii $r_2$ to $r_9$ denote the locations of thin shells, where $r_2$ to $r_9$ are $0.02, 0.03, 0.05, 0.08, 0.14, 0.22, 0.36, 0.59$ times the virial radius, 76 kpc/h, respectively.
The width of thin shells is 1/300 virial radius, about 3 cells of our computation box. The unit of time is the ground state period, which is $3.6\times 10^{-4} H_0^{-1}=5.0\times 10^6$ yrs for this $7\times 10^{10} M_\odot$ halo. Notice that the ground state (soliton) is excluded from the calculation of the halo time correlation function. Fig. (\ref{fig:corr}) shows that the correlation time of the inner halo, in between $r\approx 0.02\ r_{vir} -0.14\ r_{vir}$, is roughly the same, whereas the correlation time rapidly increases with radius.  This trend is actually expected since highly excited bound states that contribute to the outer halo have energies, thus eigen-frequencies, close to zero. 

%This property might be able to explain the formation of globular clusters. Local high density regions should survive long enough for the small scale gas to fall in. Further investigations with simulations that include baryons are needed to test this conjecture.

By defining the correlation time $\tau$ as the time when the peak correlation drops to half of its maximum value, we plot $\tau$ as a function of radius in Fig. (\ref{fig:tau}). The correlation time $\tau$ at $0.59\ r_{vir}$ in the outer halo can be nearly one order of magnitude larger than that in $r<0.14\ r_{vir}$ (inner halo), indicating relatively slow potential fluctuations in the outer halo. 

The correlation functions at inner radii are seen to still fluctuate with low-level but finite amplitudes even after 20 ground state periods in Fig.(\ref{fig:corr}), while those at the larger radii monotonically decrease to zero. These residual fluctuations arise from the fact that lowly excited states, dominant at small radius, do not have a sufficient number of states to decorrelate the fluctuations, and granules in the inner halo have long-term memory.

%???From Fig. (\ref{fig:ArtihaloDen2}), one can see that the granules size barely changes with radius. Therefore, the dynamical time of granules increases with the radius.
%We also estimate the random velocity of granules turbulent flow velocity. If we treat granules as macro-particles, the random velocity could be calculated by $v\approx 2\Delta r/\tau$, where $\Delta r$ is the typical radius of granules.??? 
%From Fig. (\ref{fig:corr}), one find that correlation time is $\tau\approx 4/3\times 2\pi\hbar/E_g\approx 7.8\times 10^{-4} H_0^{-1}$ for shell 2 to shell 6.
% *
% ^.
The thermal property of $\psi$DM is ultimately related to the granule size, as the $\psi$DM halo relies on granules to counter self-gravity. The typical granule size can be evaluated by spatial Fourier transform. The halo is divided into several thick shells, with width about 17 cells centered around the narrow shells mentioned above.  We also include a central sphere of radius $r_1$ just enclosing the central soliton. Figure (\ref{fig:PS}) shows the arbitrarily normalized spatial power spectra of the density fluctuations ($\delta$ of Eq. (\ref{eq:delta})) in these thick shells and the central soliton. The peak position of the power spectrum reflects the typical size of structure in the shell, and we see the granule size increases with the radius very mildly, changing only by a factor of $2$ over the entire halo. The average size of granules in the inner halo is about the soliton size, $r_{sol}=0.39$ kpc/h, supporting the claim made by \cite{schive_understanding_2014}. The slightly non-uniform granule size over the halo may explain why the virial temperature is smaller than $1/\beta$ of the distribution function found in Sec. (\ref{sec:result}).

The traveling speed of the granule is about $\Delta r/2\tau$, where $\Delta r\approx \pi/k_{peak}$ is the typical size of granules, and $k_{peak}$ is the corresponding wavenumber at the peak of the power spectrum (Fig. (\ref{fig:PS})). %(We call it traveling speed assuming that the granule lifetime is longer than the correlation time.)
We divide $\Delta r/2$ instead of $\Delta r$ over $\tau$ since the correlation time $\tau$ is the time when granules shift a distance roughly half of their characteristic length. On the other hand, we can define the \enquote{thermal} speed of $\psi$DM as $\hbar k_{peak}/m$ arising from the quantum pressure. Both speeds are shown in Fig. (\ref{fig:Ther_vs_Trav}) as a function of $r$.  Clearly, the \enquote{thermal} speed dominates the traveling speed over the entire halo by a large margin.   Thus if the $\psi$DM halo can be regarded as in a turbulent state, the turbulence is at most subsonic. In fact the low travelling speed of granules requires a relatively long correlation time, which in turn is simply a reflection of small energy gaps between spatially adjacent eigen-states relative to the eigen-energy.

\section{Discussions and conclusion}\label{sec:conclude}

We have successfully devised a novel method to construct self-consistent solutions of density and potential for the $\psi$DM halo where the distribution function is described by the fermionic King model. The self-consistent solutions are very stable in dynamical simulation tests. We have also examined the time correlation function for the halo granules and found that the granule coherence time can increase by one order of magnitude from the inner halo to the outer halo, despite the fact that the granule size changes little.

In this work, we also construct a Milky-Way-sized halo of $8\times 10^{11} M_{\odot}$.  This inner halo has a flatter profile than the CDM inner halo, though the outer halo appears similar. Recent modeling of observational data of Milky Way bulge with star surface brightness and kinematics does not favor the NFW profile with low concentration parameters $(c\sim 10)$ conventionally expected for the $10^{12} M_\odot$ halo \cite{portail2017chemodynamical}. This is consistent with our result that our Milky-Way-sized halo appears to agree better with a high-concentration NFW profile with $c\sim 20$, despite the fact that our inner halo significantly deviates from the NFW $-1$ power-law profile. It is known that the inner halo of CDM is built sequentially at ever increasing radius when the cosmic average density is high and minor mergers are abundant. However, $\psi$DM suppresses small galaxies and hence the galaxy assembly history is different. Moreover, CDM has a cold inner halo, but $\psi$DM has a hot inner halo characterized by the inner halo granule size.  These considerations point to an inevitable difference in the inner halo profiles of the two models.  

The fact that the granule size only changes by a factor of $2$ across the $\psi$DM halo indicates that the $\psi$DM halo has better thermal conductivity than the CDM halo. And the fact that the inner $\psi$DM halo is hot as opposed to the cold inner CDM halo is caused by the 'hot' central soliton, with which the innermost halo has thermal contact. The soliton serves as a 'heat' engine. As the soliton grows in mass along with the halo and its size is reduced, the soliton must release heat and this amount of heat is to be absorbed by the halo.  

This project is not complete without a dynamical test of the massive galaxy by simulations.  However, the granule size, in this case, is about $150$ pc and the halo size over $150$ kpc.  The entire halo needs to be resolved below $50$ pc to be able to accurately capture the granules.  This is a highly non-trivial task and will be left as a separate future work as this project continues. 

%The quiescent outer $psi$DM halo may have an interesting implication in halo star formation, given that high level of density fluctuations exist there. For the $7\times 10^10$ solar mass halo we examined, the granule coherent time there can be as long as $10^8???$ years.  A minor fraction of infalling cold gas at high redshift toward the halo center may on its way be trapped in local minima of granule potentials. For example, at half of the virial radius (30 kpc in Fig.(7)), the overdensity is $500$ times the background density and the free fall time into the local minimum is about $(\sqrt{500} H(z))^{-1}$, which amounts to $5\times 10^8$ years when $z=2$.  

\begin{acknowledgments}
S.C.Lin is grateful to Yu-Ching Shen for discussions, to Shan-Wei Lin for helping him debug and to Dr.James H.H.Chang for providing his power spectrum code. This work is supported in part by MOST of Taiwan under Grant No. MOST 103-2112-M-002-020-MY3.
\end{acknowledgments}

\appendix 
%\section{}
\section{Wigner function}\label{Wigner}

From a different perspective from Sec. (\ref{sec:DF}) , we may relate Eq. (\ref{eq:app}) to the Wigner function, a representation of the wave (or quantum) mechanical phase space distribution function analogous to $f({\bf x},{\bf p})$ of classical mechanics.  We have
\begin{align}\label{eq:dis}
\sigma_I^2\sum_{j\in I}r_j^2|\Phi_j({\bf x})|^2\approx \left|\psi_I({\bf x})\right|^2,
\end{align}
where $\psi_I({\bf x})=\sum_{j\in I}\sigma_I r_j\Phi_j({\bf x})$, since the random number $r_ir_j$ in the cross term with $i\neq j$ can be averaged to zero in the cross-term summation $\sum'_{i,j\in I}$. Now it is straightforward to show that
\begin{align}\label{eq:Wig}
&|\psi_I({\bf x})|^2=\notag\\
&\int d^3p\left[{1\over(2\pi\hbar)^3}\int d^3y \psi_I({\bf x}+{\bf y}/2)\psi^*_I({\bf x}-{\bf y}/2)e^{i{\bf p}\cdot{\bf y}/\hbar}\right],
\end{align}
where $\hbar$ is the Planck constant.
The integrand in the squared bracket on the right-hand side is nothing more than the Wigner function near the constant of motion I. 
The integration $\int d^3p$ in Eq. (\ref{eq:Wig}) simply gives the phase volume near I. Therefore, Eq. (\ref{eq:dis}) is indeed the distribution function f(I) multiplied by the spatial-dependent phase space volume near I, which we denote by $\Omega_I({\bf x})$. %The magnitude of phase-space volume $\Omega_I(x)$ changes when the value of I changes.
We can further separate the magnitude from the spatial dependence, i.e., $\Omega_I({\bf x})=g(I)h_I^2({\bf x})$. Here, $h^2_I({\bf x})$ is the weighted average of all $|\Phi_j({\bf x})|^2$ within I, thus $h_I^2({\bf x})=\sum_{j\in I} r_j^2|\Phi_j({\bf x})|^2/\sum_{j\in I}r^2_j$, and hence $\int h^2_I({\bf x})d^3x=1$. At the end, we arrive at
\begin{align}\label{eq:rhoav}
<\rho>({\bf x})=\sum_If(I)g(I)h^2_I({\bf x}).
\end{align}

\section{Distribution functions of self-gravitating systems}\label{sec:CDF}

In this work, we consider a few well-known distribution functions, which are either a function of energy or a function of both energy and angular momentum.  We briefly describe these distribution functions for references in the main text.

\subsubsection{Models as functions of energy}\label{sec:FK}
The first model is the King model \cite{king_structure_1966}, or the lowered isothermal model, which behaves like Maxwell-Boltzmann distribution when energy is far below the negative escape energy. The density profile does not extend to infinity. That is, the system is truncated at a certain escape energy to have a finite mass. The distribution function of the King model is
\begin{align}
f_{\text{King}}=
\begin{cases}
A(e^{-\beta(E-E_c)}-1),&\text{if } E\leq E_c\\
0, &\text{otherwise}
\end{cases}
\end{align}
 where $E_c$ is the escape energy, and $\beta$ can be interpreted as inverse temperature.
 
The second model is the fermionic King model, which is proposed by Ruffini and Stella \cite{ruffini_semi-degenerate_1983}, and can be derived from classical kinetic theory \cite{chavanis_coarse-grained_1998}. This model is motivated by the Lynden-Bell's distribution for collisionless particles which in some simplified cases is described by Fermi-Dirac distribution. The fermionic King model differs from the King model only by dividing a Fermi-Dirac factor. That is,
\begin{equation}\label{eq:6}
f_{\text{FK}}=
\begin{cases}
A\frac{e^{-\beta(E-E_c)}-1}{e^{-\beta(E-E_c-\mu)}+1},&\text{if } E\leq E_c\\
0, &\text{otherwise}
\end{cases}
\end{equation}
 where $\mu$ is the chemical potential. If $\mu\rightarrow -\infty$, the fermionic King model reduces to the King model. The amplitudes of lower excited states in the model are suppressed in the presence of chemical potential.

\subsubsection{Models as functions of energy and angular momentum}

Spherical-symmetric self-gravitating system can have anisotropic velocity dispersion if the distribution function is a function of both energy and orbital angular momentum $L^2$. That is,
\begin{align}
f=f(E,L^2)
\end{align}
We consider the Osipkov-Merritt model, which is generated by replacing the argument of the distribution function from energy E to Q, where Q is defined as
\begin{align}\label{eq:Q}
Q \equiv -E-\frac{L^2}{2r_a^2},
\end{align}
where $r_a$ is a constant scale radius. The velocity dispersion inside $r_a$ is isotropic, whereas it becomes radially biased in the region outside $r_a$ \citep{binney_galactic_2011}. The Osipkiv-Merritt version of the King model is
\begin{align}
f_{\text{OMK}}=
\begin{cases}
A(e^{-\beta(Q-Q_c)}-1),&\text{if } Q\leq Q_c\\
0. &\text{otherwise}
\end{cases}
\end{align}
\iffalse%%%%%%%%%%%%%%comment%%%%%%%%%%%
We also consider the Michie model, which takes the form
\begin{align}\label{eq:10}
f_M=
\begin{cases}
Ae^{-\beta L^2/(2r_a^2)}(e^{-\beta(E-E_c)}-1),&\text{if } E\leq E_c\\
0, &\text{otherwise}
\end{cases}
\end{align}
$r_a$ again represent the scale radius. In the limit $r_a\rightarrow\infty$ equation (\ref{eq:10}) reduce to the King model, and if
\begin{align}
\beta(E-E_c)\gg 1
\end{align}
\begin{align}
f\approx Ae^{-\beta L^2/(2r_a^2)}e^{-\beta(E-E_c)}=Ae^{Q},
\end{align}
which is one of the Osipkiv-Merritt models because it only depends on $Q$.
\fi%%%%%end comment%%%%%

\bibliography{thesis_editing} 
\bibliographystyle{unsrt}
\end{document}